\documentclass[twocolumn]{webofc} %Swapped to 2 column because we were over 4 pages in single column format 

\usepackage[varg]{txfonts}   % Web of Conferences font
\usepackage{hyperref}
\usepackage{url}
\usepackage{graphicx}
\usepackage{subcaption}

\hypersetup{colorlinks=true,citecolor=blue,urlcolor=blue,linkcolor=blue}
\begin{document}
\title{Pressure Waves During Granular Flows in Varying Gravity Environments}
%
% subtitle is optionnal
%
%%%\subtitle{Do you have a subtitle?\\ If so, write it here}

\author{\firstname{Abigail} \lastname{Tadlock}\inst{1}\fnsep\thanks{Current: Boston University, Astronomy Department} \and
        \firstname{Lori} \lastname{McCabe}\inst{1}\fnsep\thanks{Current: Colgate University, Physics \& Astronomy Department} \and
        \firstname{Kerstin} \lastname{Nordstrom}\inst{1}\fnsep\thanks{\email{knordstr@mtholyoke.edu}}
}

\institute{Department of Physics \& Astronomy, Mount Holyoke College, South Hadley, MA, USA}

\abstract{We present results of LAMMPS Molecular Dynamics simulations of 2D gravity-driven flows of 30,000 soft uniform spheres through a vertical silo. We vary the gravitational field ($G$), elastic modulus of the particles ($E$), and silo outlet diameter ($D$). We present results on upwards pressure waves observed in the system. We compare our results with previous work on granular acoustics. Despite the typical particle speed being substantially less than the measured wave speed, we posit these are nonlinear shock waves, as observed in other systems near jamming. We demonstrate that the wave speeds in all systems appear to follow a power law that is distinct from linear wave expectations. }
\maketitle

\section{Introduction}
\label{intro}
The study of granular discharge flow from a silo is of great importance to many industries. It is also a simple system with inherent complexity, with non-uniform stresses and particle movements within the same system that may be quasistatic (``slow'') through inertial (``fast''). The mass discharge rate in granular silos is usually adequately described by the empirical Beverloo equation \cite{beverloo_flow_1961}, which depends on the ratio of outlet diameter ($D$) to particle diameter ($p_d$). The equation also includes gravity as a factor of $\sqrt{G}$. Previous experimental work has tried to investigate this scaling
\cite{mathews_model_2016} using a large centrifuge, but had limited dynamic range. Understanding the role of gravity in basic granular processes is important, especially with the advent of increased planetary/asteroid surface exploration. 

Pulsatile behaviors have been observed in industrial silos for years, and sometimes create unsettling shock waves at frequencies humans can hear \cite{roberts_flow_2002}. The mechanism of this is stick-slip at the wall, and the actual discharge rate with time may be affected. In contrast, we observe more subtle pressure waves that do not significantly impact the discharge rate itself, but are nonetheless present, suggesting these waves are a universal feature of silo flow. We find dependencies on elastic modulus and gravity, and compare to other work on granular acoustics. 
% \begin{equation}
%     W = C\rho_b\sqrt{g}(D_0 - kp_d)^{3/2}
%     \label{eq:beverloo}
% \end{equation}

\section{Simulations}
\label{LAMMPS}
Simulating granular flow allows us to easily vary conditions that are difficult to experimentally alter, such as gravity. We use the \it granular \rm package of LAMMPS (Large-scale Atomic/Molecular Massively Parallel Simulator) to create our system \cite{thompson_lammps_2022}. This package uses the Hertz-Mindlin model of particle interaction. Some parameters we keep constant for all simulations, such as the coefficient of restitution of 0.66, coefficient of static friction of 0.5, and a Poisson's ratio of $2/7$, similar to other work using this model \cite{silbert_granular_2001}.

%Remove hertizan specifics for space? 
% This model assumes that the force felt by two particles (particle $i$ and particle $j$) of effective mass, $m_{eff}$, with overlap distance, $\delta$, is nonlinear, proportional to the overlap area. The Hertzian contact force for a monodisperse system with radius $R$ is as follows:
% \begin{equation}
%      \vec{F}_{hz} = \sqrt{\frac{\delta R}{2}} (k_n \delta \hat{n}_{ij} - m_{eff} \gamma_n \vec{v}_n - k_t \Delta \vec{s}_t - m_{eff} \gamma_t \vec{v}_t)
% \end{equation}
% where $k_n$ and $k_t$ are nonlinear spring constants with the units of force/area in the normal and tangential directions, respectively, and depend on Young's modulus, $E$. $\hat{n}_{ij}$ is the unit vector connecting the center of the two particles. $\gamma_n$ and $\gamma_t$ are viscoelastic damping constants. $\Delta \vec{s}_t$ contains information about the tangential displacement of the particles over the period of contact. $\vec{v}_t$ and $\vec{v}_n$ are the tangential and normal velocities.

We simulate a quasi-2D system of monodisperse spheres in a flat-bottomed silo that is 63.1 particle diameters ($p_d$) wide x 700 $p_d$ tall x 1.1 $p_d$ deep. Quasi-2D systems are frequently used in silo flow research and serve as analogues to 3D systems \cite{hong_clogging_2017, hong_clogging_2022, endo_obstacle-shape_2017}. The $p_d$ = 1 m, and the density, $\rho$, is 1 kg/m$^3$. Our particles are therefore unphysically large, but these values may be thought of as simulation units. The simulation uses timesteps of 0.0001s, with data writeout every 1000 timesteps (every 0.1 seconds). We systematically vary three elements of the simulation: gravity, elastic modulus, and outlet diameter. Gravity, $G$, is varied between 0.1 and 20 times Earth's gravity, $g$ = 9.8 m/s$^2$, the elastic modulus $E$ is varied between $10^4$-$10^7$ Pa, and the outlet diameter is varied between 6$p_d$ and 20$p_d$. We define a dimensionless constant $\Gamma$ in equation \ref{eqn:gamma} to navigate within this phase space and compare different simulations.
 \begin{equation}
     \Gamma = \frac{\rho G p_d}{E} \propto \frac{G}{E}
    \label{eqn:gamma}
 \end{equation}
 
\begin{figure}[h]
    \centering
    \includegraphics[width=0.40\textwidth]{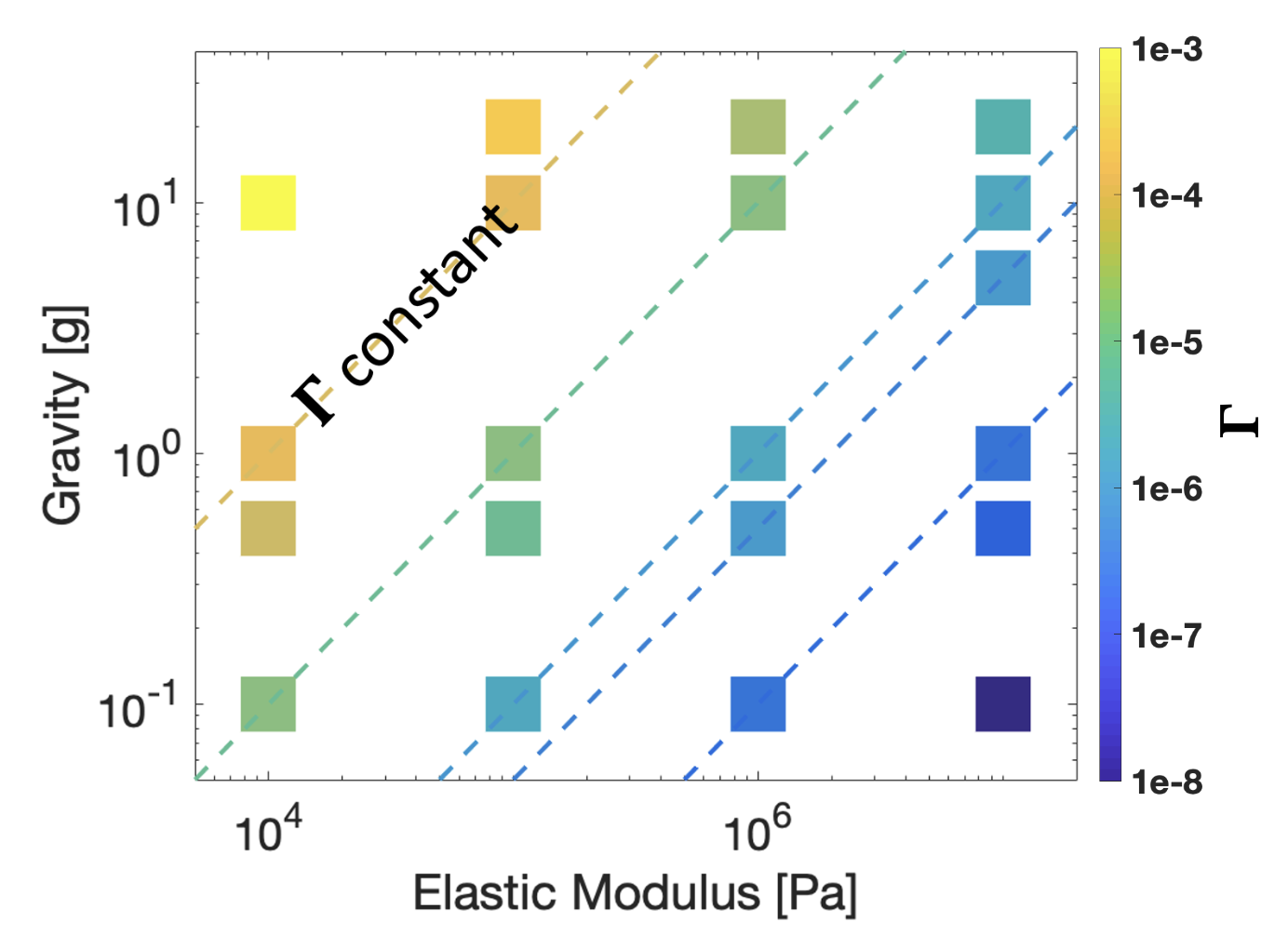}
    \caption{Phase space of $\Gamma$ as a function of gravity ($G$) and elastic modulus ($E$). In our simulations, $\rho$ and $p_d$ are constant, so the only varibles affecting $\Gamma$ are $G$ and $E$.} 
    \label{fig:phasespace}
\end{figure}

Physically, $\Gamma$ (or powers of it) can be interpreted as a ratio between elastic and gravitational timescales, speeds, forces, etc. In our simulations, since the particle diameter $p_d$ and the density $\rho$ are both 1 in their respective units, $\Gamma$ is numerically equivalent to $G/E$. Throughout our simulation phase space, many simulations have the same values of $\Gamma$ and we observe some similarities in behavior in these systems' mass flow rates and other characteristic measurements, although that is not the focus of this paper. Figure \ref{fig:phasespace} shows our phase space with dashed lines denoting simulations that have the same values of $\Gamma$.

\begin{figure}
    \centering
    \includegraphics[width=0.25\textwidth]{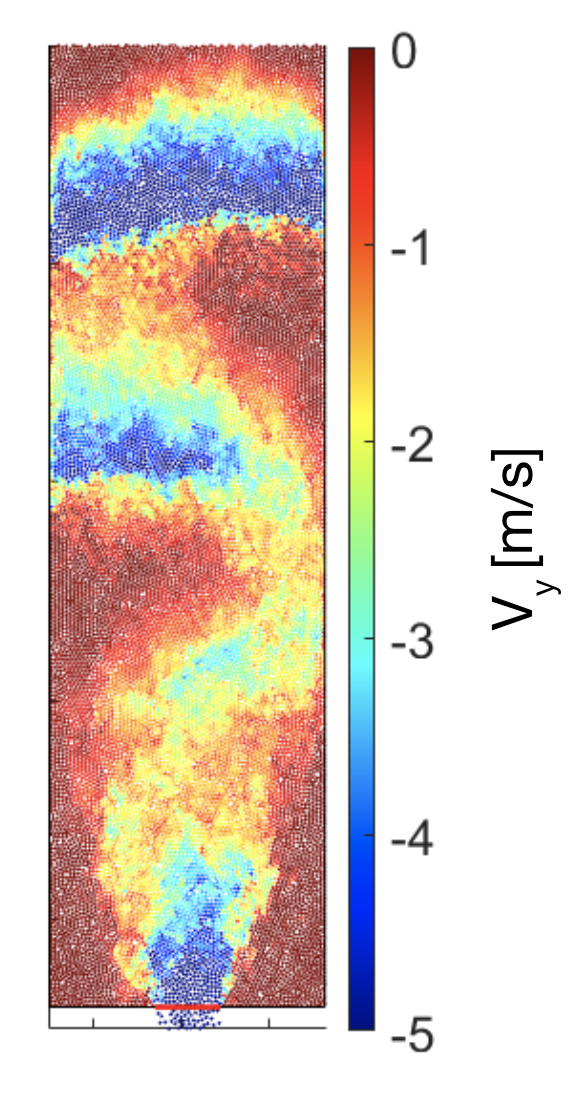}
    \caption{Pseudoimage of one  simulation ($G = 1$ g, $D_0 = 15$ p$_d$, $E = 10^4$ Pa) at a particular time. The colorscale is instantaneous vertical velocity. The dark blue ``bubbles" are the wavefronts. They travel upwards while the constituent particles are moving down at high velocity. The wavesfront also have a lower $\phi$ than other regions.}
    \label{fig:PressureWaveEx}
\end{figure}

We use the LAMMPS {\tt create\_atoms} command to insert particles into the silo. All of the particles are created at once, located at random positions with random velocities within the silo, and allowed to settle for an extended period of time such that the kinetic energy is minimized. The random initial positions and velocity depend on a seed specified in the code, which we did not vary. As a result, the initial packings for particular $G$/$E$ combinations are identical.  After the packing has settled, we remove a ``plug" at the outlet instantaneously (in a single data time-step) and allow the particles to flow with no further modifications to the simulation. Typical values of bulk volume fraction ($\phi$) at $1g$  range from 0.528 (stiffest) to 0.703 (softest), and coordination numbers range from 4.27 to 5.45, for the same range. The corresponding particle deformation ($\delta/R$) is around 10 percent (softest), and smaller for stiffer systems.  Also, while not the focus of this paper, we also observe lower volume fractions near the outlet, similar to previous work such as \cite{endo_obstacle-shape_2017}.

\begin{figure}
    \centering
    \includegraphics[width = 0.8 \linewidth]{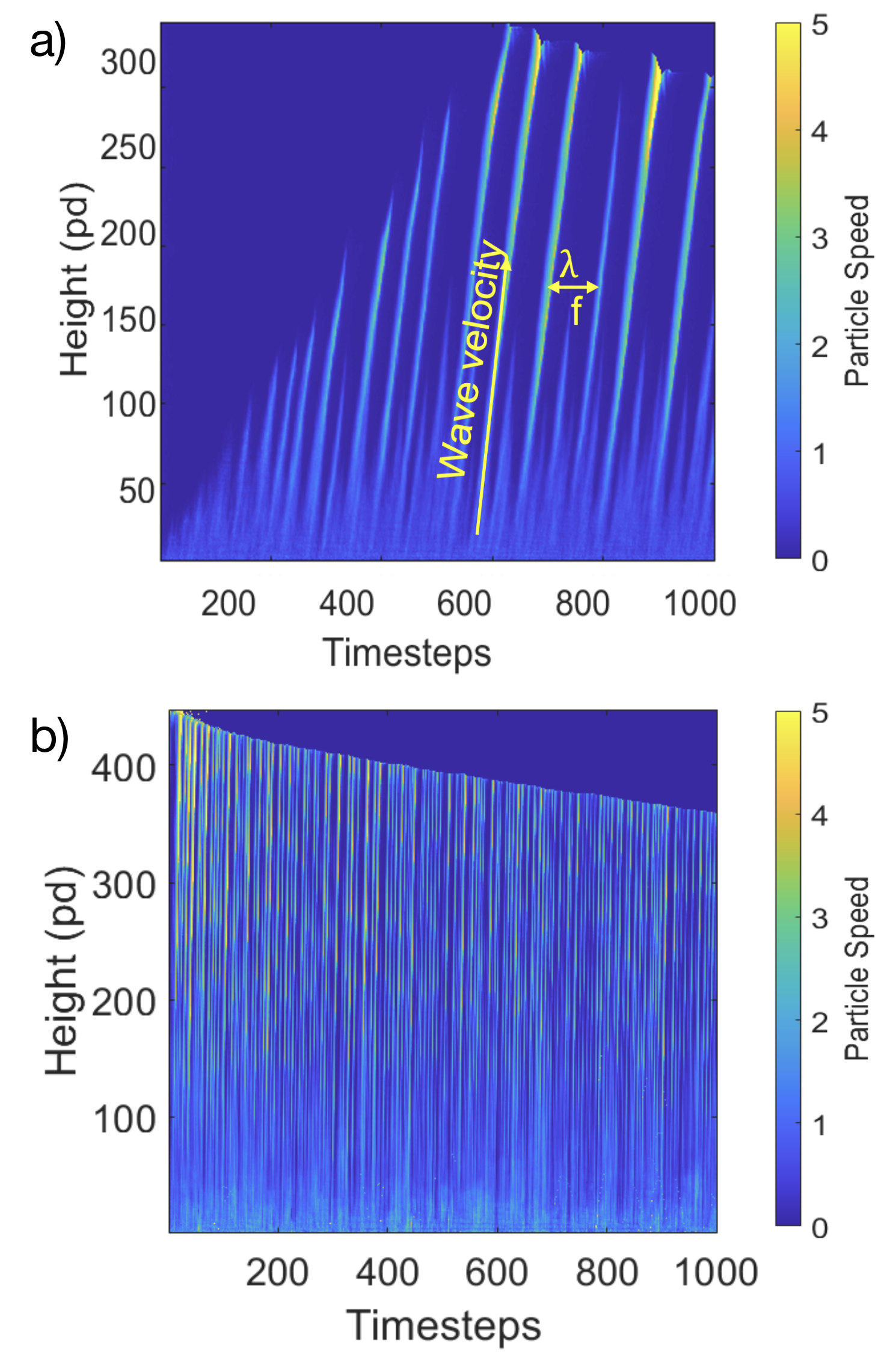}
    \caption{Spacetime diagrams for two systems with the same gravity, $G = 1$ g, and outlet, $D_0 = 9$ p$_d$, but different values of particle stiffness, (a) $E = 10^4 $Pa (b) $E = 10^7$ Pa. The slope of the line is the wave velocity and the separation gives both frequency and wavelength, as shown in (a). %cut from caption: The lower stiffness particles have less frequent waves and lower wave speed, while the harder particle systems have frequent and fast waves.
    }
    \label{fig:WaveFreqComp}
    
\end{figure}

\section{Pressure Waves}
\label{PressureWaves}

We see pressure waves develop in all systems, illustrated in Figure \ref{fig:PressureWaveEx}. These waves travel upward in the silo, opposite the gravitational flow of the particles. Generally, a softer system will have less frequent waves and a slower wave speed than a stiffer system with the same value of $G$. The softer systems also have more attenuation. We do not see any dependence on outlet size. In some simulations, the initial wave pulse was clear to the eye: when a void formed near the outlet, particles along the sides would rush into the empty space, catalyzing an upward rarefaction wave. We have separately verified that these pulses have a slightly lower particle density than the packing structure of the surrounding particles.

\begin{figure}[h!]
    \centering
    \includegraphics[width=\linewidth]{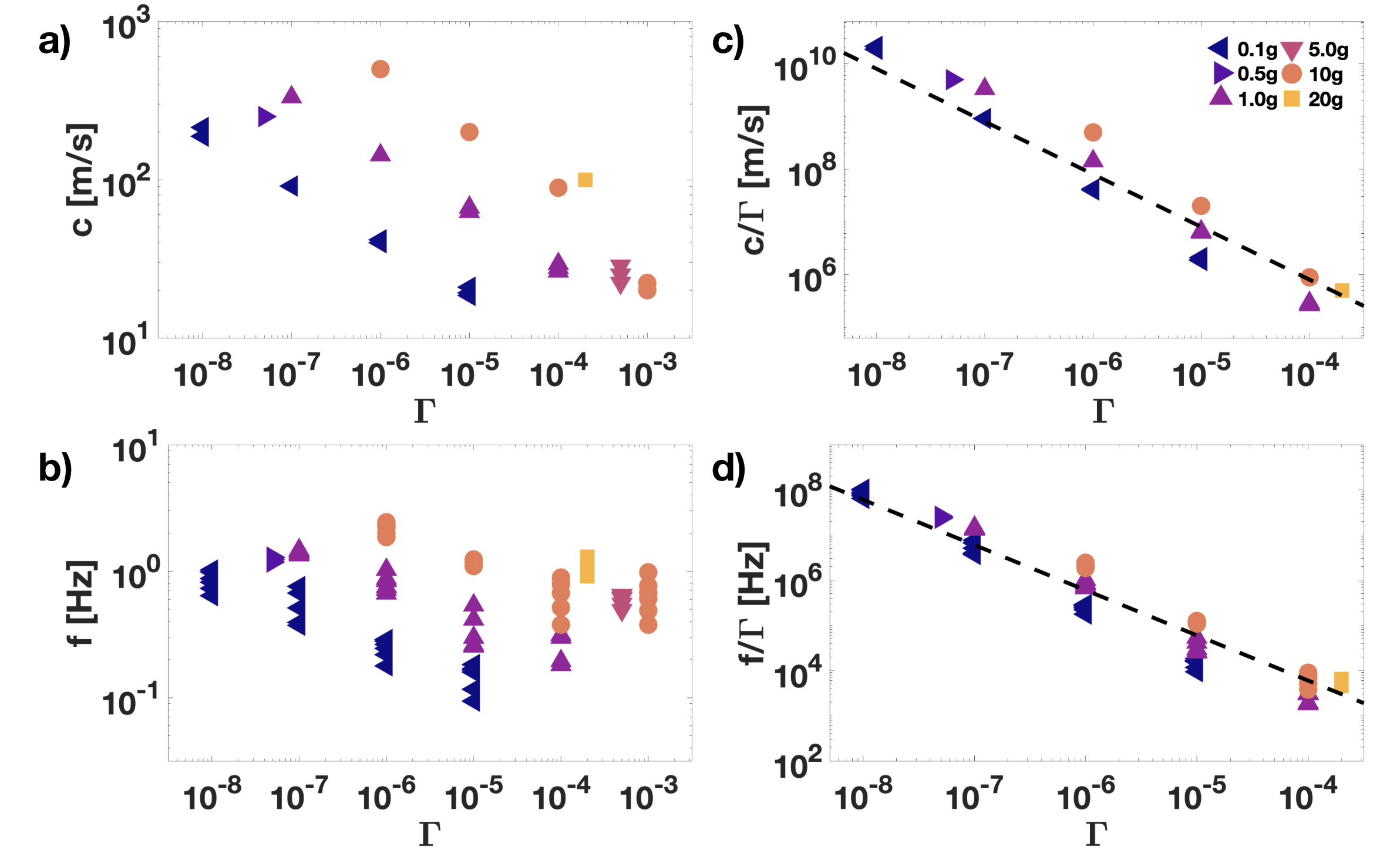}
    \caption{The raw (a) wave speed and (b) frequency of each system, and the normalized by $\Gamma$ (c) wave speed and (d) wave frequency, fitted with a -1 power law fit. We note the velocity had multiple datapoints of the same value, so they overlap. The dataset is the same size in all subfigures.}
    \label{fig:waveanalysis}
\end{figure}

\begin{figure}[h]
    \centering
    \includegraphics[width=0.31\textwidth]{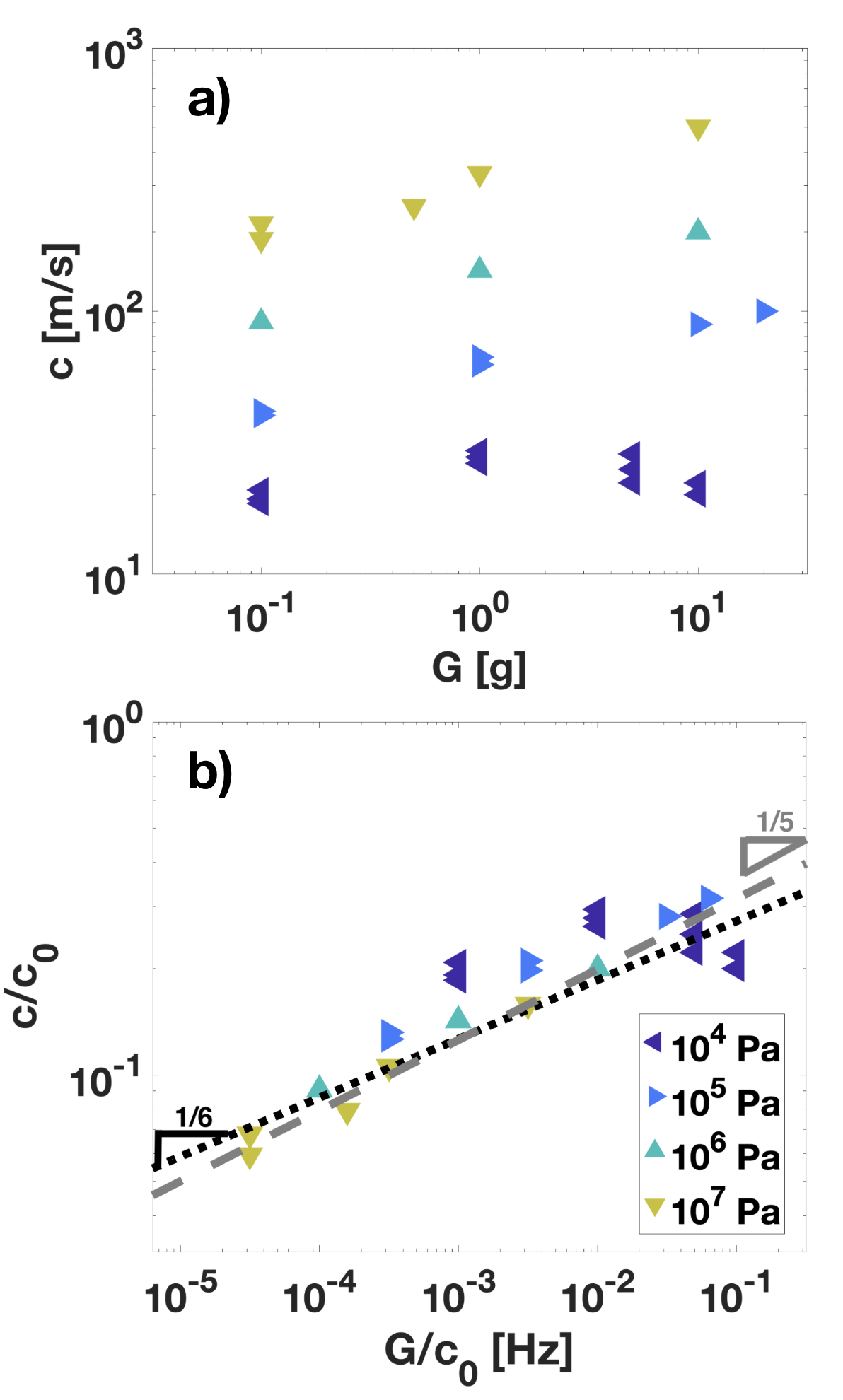}
    \caption{Wave speed versus gravity (a) raw data and (b) scaled by the expected speed of sound for a continuous medium, $c_0$. The collapsed data has two potential power law comparisons, of $1/5$ (dashed) and $1/6$ (dotted).}
    \label{fig:Vgcompare}
\end{figure}

To analyze these pressure waves, we generate spacetime diagrams by averaging the instantaneous velocity of all particles at a given height. This creates a vertical stack of averages throughout the height of the silo per timestep, and we plot the sequence of these slices against time on the $x$-axis, shown in Figure \ref{fig:WaveFreqComp}. The pressure waves appear as diagonal traces on the spacetime diagram. The softer particles in Figure \ref{fig:WaveFreqComp}a have less frequent waves and lower wave speed, while the stiffer particle systems in Figure \ref{fig:WaveFreqComp}b have frequent and fast waves. We also calculate the wave velocity, $c$, the wavelength (not shown), $\lambda$, and frequency, $f$, in each system as annotated in  \ref{fig:WaveFreqComp}a. The wave speed is on the order of 100 m/s, while the particles move on the order of 5 m/s individually ($5 \times 10^{-4}$ $p_d$ per timestep). In the bulk, the inertial number is $I <10^{-3}$ suggesting a quasistatic treatment of the granular material is appropriate. Near the outlet, the flow becomes dense ($10^{-3}<I < 10^{-1}$).

Wave speed $c$ and frequency $f$ are plotted for all systems in Figure \ref{fig:waveanalysis}a-b. For longitudinal wave transmission through a continuous medium, the wave speed is $c_0 = \sqrt{E/\rho}$, giving a value for our softest systems of $100 $ m/s and the stiffest systems $3100 $ m/s. Our wave speeds are consistently lower than the continuous medium estimate, which is to be expected. Furthermore, with the relationship of $E$ and $\Gamma$ in Equation \ref{eqn:gamma}, a zeroth-order approach would suggest a wave speed dependence of $\Gamma^{-1/2}$, but the unscaled data in Figure \ref{fig:waveanalysis}a-b shows a $\Gamma^{-1}$ dependence. We also rescale the data by $\Gamma$ in Figure \ref{fig:waveanalysis}c-d, and the data collapses with an additional power law dependence of $\Gamma^{-1}$, suggesting the physics is universal. 

Of course, granular materials are not continuous media. The elastic modulus is roughly predicted by effective medium theory, but there are discrepancies observed. For Hertzian contacts, the prediction for linear waves is that $c \propto P^{1/6}$ \cite{owens_sound_2011,coste_validity_1999}, where $P$ is the confining pressure. Discrepancies observed are perhaps due to nonaffine motions, linear force networks, or other heterogeneities (e.g. stresses, packings) \cite{somfai_elastic_2005}.

In \cite{owens_sound_2011}, the average pressure was taken to be the pressure near the bottom of the cell. This pressure will exhibit a Janssen saturation (will not depend on fill height, unlike a fluid system), but should still scale with $G$. We have a similar geometry here and thus directly vary the confining pressure in our simulations by varying $G$. For linear wave propagation we thus expect a scaling of wave speed $c \sim G^{1/6}$.  However, we are also varying the ``disturbance'' pressure in the same manner, as the wave pulse appears to be a consequence of particle free-fall into the exit region. Future work will look into the mechanics of this formation in more detail, for now we focus on its effects. 

Judging by the relative speed of the waves and the speeds within these materials, this does not appear to be a shock-like phenomenon at first glance, as our particle speed is much less than $c$. However, materials near the jamming point may experience these nonlinearities regardless, as seen in work such as \cite{gomez_shocks_2012, gomez_uniform_2012, van_den_wildenberg_shock_2013}. The basic argument is that, near jamming ($\phi_j$), the rigidity (and pressure) vanishes as powers of $\phi-\phi_j$, so that the effective speed of sound is actually very low. Thus any disturbance can produce a supersonic shock in the material \cite{gomez_shocks_2012}. Near the outlet in all cases, the packing is inherently near $\phi_j$, due to the random packing and marginal stability. Thus the ``impact'' may act as a shock traveling upwards the system. We do not know $\phi_j$ in our systems, and moreover estimating $\phi-\phi_j$ would be more fraught due to experimental uncertainties. 

Previous work has looked at shocks due to impacts, such as \cite{van_den_wildenberg_shock_2013, clark_nonlinear_2015}, and suggest alternate techniques. In particular, \cite{clark_nonlinear_2015} derives an equation useful for our analysis: 

 \begin{equation}
   \frac{c}{c_0}\propto \left(\frac{v_G}{c_0}\right)^{(\alpha-1)/(\alpha+1)} 
    \label{eqn:shock}
 \end{equation}

\noindent In this equation, $v_G$ is the characteristic grain speed based on free-fall, and $\alpha$ depends on the (normal) force law between two particles in contact: $F\propto \delta^{\alpha}$, where $\delta$ is the compression. For Hertzian particles $\alpha = 3/2$. Thus, we might expect our data to follow Eq. 2  with a power of 1/5. Our characteristic velocity should scale with $G$, as a particle in free-fall.  

The unscaled data is plotted in Fig. \ref{fig:Vgcompare}a as $c$ vs $G$. We observe a promising trend with $G$, aside from 2 conditions at high gravity and low elastic modulus (dark blue). Scaling the data is shown in Fig. \ref{fig:Vgcompare}b. The data collapses well, indicating that Eq. 2 is promising. We overlay two power laws, corresponding to 1/6 (linear) scaling and 1/5 (shock) scaling. We do not formally claim one is better, as fitting such power laws is tricky, although 1/5 appears by eye to be more promising with our data. If we eliminate the two conditions that seem to deviate, the steeper 1/5 law becomes more compelling. 

However, this doesn't quite address whether the confining pressure or the ``shock'' pressure is really at play, as both will scale proportionally with $G$. It is worth mentioning that  \cite{van_den_wildenberg_shock_2013} was able to independently tune both pressures in impact experiments. Confining pressure only had a weak effect, if at all. Impact pressure had no effect on wave speed at ``low'' pressure, but for high impact pressure had power law scaling $c\sim P_{impact}$. They saw $1/6$ scaling, but others have found $1/5$ or even $1/4$ scaling with similar stimuli \cite{gomez_shocks_2012,gomez_uniform_2012,somfai_elastic_2005, clark_nonlinear_2015}. 

Thus, we conclude that we observe shock waves in our system, precipitated by void formation and resultant filling (our disturbance), as mass flows out of the outlet region. What sets the size and frequency of the wave pulses is to be further investigated, as well as the attenuation. While these appear to be more collective than transmission along force chains as in \cite{clark_nonlinear_2015,owens_sound_2011}, we have not looked into the local forces in more detail, as LAMMPS only outputs net forces on particles. A workaround with the current data would be to look into nonaffine motions more specifically.

In all simulations, the value of $\Gamma$ was quite low, indicating that the (particle material) elastic speed is high compared to the gravitational speeds. Our largest values of $\Gamma$ correspond to the squishiest systems, and those seem to deviate from other results, suggesting some critical $\Gamma$ for which these gravitational/elastic truly compete. 

This raises two more open questions of note for highly squishy systems. In \cite{clark_nonlinear_2015}, a deviation from 1/5 scaling was also seen at high particle speed ($G$ in our case) and softer particles. However, they saw an upward deviation while we see a downward deviation. They hypothesize the deviation is due to more contacts being loaded as particles deform more extremely. We agree this makes sense, but nonetheless see a different deviation. This could be due to differences in force law (their particles were not Hertzian) or the viscoelastic properties of their particles. Lastly, \cite{van_den_wildenberg_shock_2013} claim that for ``more elastic particles, or for microgravity," shock waves may be observed as dissipation-free. We generally have soft particles and many trials with low gravity. We always saw attenuation in our trials (we did have friction and collisional losses), but this presents an obvious avenue for future simulation work.

\section{Acknowledgments}
This work was supported by United States National Science Foundation grant DMR-1846991.

% BibTeX or Biber users please use (the style is already called in the class, ensure that the "woc.bst" style is in your local directory)
% \bibliography{your_bib_file} % Replace "your_bib_file" with the actual name of your .bib file
%
\bibliography{references}
\end{document}